\documentclass[aps,prd,amsmath,amssymb,12pt]{revtex4}
\usepackage{graphicx}% Include figure files
\usepackage{color}
\usepackage{bm}% bold math
\def\journal#1#2#3#4{{\em #1}, {\bf #2} #3 (#4)}
\newcommand{\be}{\begin{equation}}
\newcommand{\ee}{\end{equation}}
\newcommand{\bea}{\begin{eqnarray}}
\newcommand{\eea}{\end{eqnarray}}
\newcommand{\hf}{\frac12}
\newcommand{\nn}{\nonumber\\}
\def\eq#1{(\ref{#1})}

\def\hphi{{\hat\phi}}

\def\hD{{\hat D}}
\def\mr#1{{\mathrm{#1}}}
\def\ord#1{{\cal O}\left(#1\right)}

\def\vv#1{{\bm{#1}}}

\begin{document}
\title{Boost invariant regulator for field theories}
\author{Janos Polonyi}
\affiliation{Strasbourg University, High Energy Theory Group, CNRS-IPHC,23 rue du Loess, BP28 67037 Strasbourg Cedex 2 France}
\date{\today}

\begin{abstract}
It is shown that by imposing the relativistic symmetries on the cutoff in field theories one rules out all known non-perturbative regulators except the point splitting. The relativistic cutoff dynamics is non-local in time and thereby unstable, bringing the very existence of relativistic field theory into question. A stable, relativistic regulator is proposed for a scalar field theory model and its semi-classical stability is shown numerically.
\end{abstract}
\maketitle

\maketitle

\section{Introduction}
The continuously distributed degrees of freedom of a field theory lead to mathematically ill defined, infinite quantities, interpreted as inappropriate extrapolations from observed to not yet explored scale regimes. The problem is avoided by introducing a UV regulator, a cutoff in the theory. The regulated theory describes the lower lying, observed particle modes, coupled to the non-resolved UV environment and thereby defines an open system. Although the regulator is not unique, it takes into account the undetected modes in an approximate manner and reflects our ignorance, but the physical predictions within the observed scale regime, far from the cutoff scale, are supposed to be universal, i.e., independent of the particular details of the regulator. The regularization should retain some general features, expected to reign even within the not yet explored scale regimes. The properties we are mainly concerned with here are the symmetry with respect to boosts and the stability. We regard this problem in a non-perturbative setting, i.e., by defining the cutoff-dependent terms of the bare action which render the path integral finite rather than regularizing the Feynman diagrams. 

The regularization can be realized on two different levels: Green's functions are distributions, generalized functions, hence they may have certain singular structure. For instance the free Green's functions diverge when two of their arguments approach each other. One can introduce a regulator and  approach the free Green's function by the help of finite functions. Once the free theory is regulated the inclusion of the interactions can not generate divergences, c.f. lattice regularization where the theory is placed on a space-time lattice \cite{wilson}, the lattice spacing playing the role of the cutoff. One may realize a less ambitious regularization, by leaving the singularity structure of the free theory unchanged and removing only the singularities, generated by the interactions. This can be done by point pslitting, i.e. by smearing the the interaction vertices without modifying the free Green's functions \cite{psplitting,psplittinggt}.

The starting point of this work is the observation that the known non-perturbative regulators of the free Green's function break the boost symmetry and the regularization of the divergences, generated by the interactions, renders the dynamics unstable. The difficulty of regularizing the free theory stems from the infinite volume of the external, space-time symmetry group, owing to the boosts. Our main result is a modified point-splitting regularization scheme with Lorentz boost invariance and stable dynamics. While the usual regulators suppress modes according to the external space quantum numbers, energy or momentum, the modified point-splitting regulator produces a stable dynamics by suppressing the modes in the internal space, in the space of field amplitude, as well. 

Our argument starts in section \ref{regs} with a list of the known regularization schemes, followed by a brief survey of the ways two important requirements of regulators, the preservation of boost symmetry and stability are handled. The former is the subject of section \ref{symms} and the stability of cutoff theories is discussed in section \ref{dscs}. Section \ref{pointspls} contains the description of a point splitting regularization which preserves boost invariance and causality. The extension of such a regulator to the internal space with stability is the subject of section \ref{insspss}.

\section{Regulators in field theories}\label{regs}
The UV singular structure of the free Green's function arise from the unbounded spectrum of the momentum operator of the first quantized formalism and we have three different strategies at our disposal to regulate them. (a) The {\em dispersion relation is modified} and thereby the free propagator is suppressed in the UV domain by adding some higher powers of the space-time derivatives to the local quadratic part of the action, e.g., the Pauli-Villars method \cite{pv}, the use of the proper time \cite{propt} and the lattice regularization \cite{wilson}. (b) One performs a {\em Wick rotation} and continues analytically the Fourier integrals to Euclidean space-time where either an $O(d+1)$ invariant sharp cutoff is employed or a further analytic extension is made, such as the dimensional \cite{dimreg} and the $\xi$ \cite{xi} or $\zeta$-function \cite{zeta} regularizations. (c) The {\em space of states is restricted} by removing states from the Fock-space beyond a certain momentum or energy by the help of a sharp cutoff. 

If the singularities of the free theory are retained then the regularization of the UV singularities caused by the contact interactions between elementary excitations, local interaction vertices, can be achieved by {\em point splitting}, by smearing the interaction vertices \cite{psplitting,psplittinggt}. Note also that any regulator introduces a temporal or a spatial non-locality in the action at the scale of the cutoff. Furthermore it is shown in section \ref{pointspls} that the class (a), the modification of the kinetic energy, can be represented as a point splitting for the free action and a trivial rescaling of the Green's functions. 

The regularization should preserve certain features of the naive, unregulated theory, of which we single out the boost symmetry and the stability.

\section{Boost symmetry}\label{symms}
The regulator should preserve the symmetries of the observed physics. If this is not possible then the damage, made by the cutoff is different for global and local symmetries. The breakdown of a local symmetry can not be recovered \cite{elitzur}, moreover such a symmetry breaking renders a gauge theory non-renormalizable, \cite{preskill}. A global symmetry, broken by the regulator at the cutoff scale, may be restored at the finite observation scales in the renormalized theory. The recovery of the Euclidean external symmetry in the continuum limit of lattice gauge theories has indeed been checked numerically \cite{lorrec}. 

However there are physical phenomenas which are influenced by the violation of the space-time symmetries, whatever far in the UV regime do they occur. These phenomenas are denoted by a rather unfortunate name, anomalies. The violation of the space-time symmetry at the lattice spacing do not decouple from the chiral anomaly \cite{chanomaly} in the continuum limit \cite{celmaster}. Note that the radiation reaction force of a point charge is another anomaly effect in the sense that it arises from the modes of the electromagnetic fields around the cutoff scale, whatever short distance may it be \cite{lorentzsens} and the derivation of the Abraham-Lorentz force requires a manifestly Lorentz invariant regulator. 

The Lorentz symmetry at the cutoff scale remains important beyond the domain of anomalies because all theories we know are effective and apply down to a minimal distance only, to their UV cutoff. Effective theories need manifestly Lorentz invariant regulator, too. If gravity, a gauge theory of the external symmetry, needs quantization then that can only be carried out with manifestly Lorentz invariant regulator, 

The preservation of the boost symmetry is a specially difficult question because it is related to the very origin of the UV divergences. In fact, these divergences arise from the infinitely many momentum value a free particle can assume and any attempt to limit these values in the regularization procedure goes against the boost invariance. 

There is a peculiar relation between the boost and the gauge invariance in non-relativistic theories. The Galilean boost, $\vv{x}\to\vv{x}-t\vv{v}$, shifts the momentum, $\vv{p}\to\vv{p}-m\vv{v}$, and can be achieved by a gauge transformation where the angle of the gauge transformation is a linear function of the space coordinates. In other words, a Galilean boost invariant regulator requires gauge invariance, e.g., the usual sharp cutoff in momentum space breaks both the boost and the gauge invariance. However gauge invariance alone is not sufficient for the Galilean symmetry, the regularization by a spatial lattice is gauge invariant but breaks translation and rotation symmetries. The Lorentz boost rescales the momentum and can not be generated by an appropriate gauge transformation thus Lorentz boost and gauge invariance are not related.

A more complete picture of this problem is offered by the loop-integral,
\be\label{boinint}
I=\int\frac{d^{d+1}p}{(2\pi)^{d+1}}f(p),
\ee
in $d$-dimensional space with manifestly boost invariant integral measure and integrand, i.e., $f(p^0,\vv{p})=F_r(p^{02}-\vv{p}^2)$, and  $f(p^0,\vv{p})=f(p^0,\vv{p}-m\vv{v})=F_{nr}(p^0)$ in the relativistic and the non-relativistic cases, respectively. This integral is infinite, owing to the infinite volume of the external symmetry group. This is obvious in the non-relativistic case, and can be seen for a relativistic integral by writing it in the form
\be
I=\int ds^2F(s^2)\Sigma(s^2),
\ee
where 
\be\label{inv}
\Sigma(s^2)=\int\frac{d^{d+1}p}{(2\pi)^{d+1}}\delta(p^2-s^2)
\ee
denotes the area of an equi-integrand surface. The sharp cutoff in momentum space, the restriction of the integration within the region $\vv{p}^2<\Lambda^2$, yields $\Sigma(s^2)\sim\Lambda^{d-1}$ for $d>1$ and $\Sigma(s^2)\sim\ln\Lambda$ for $d=1$ according to the simple power counting. Hence the volume of any regions, bounded by Lorentz invariant quantities, is necessarily infinite. This divergence is less strong than in the non-relativistic case due to the non-linear dependence on the boost velocity but is still strong enough to plague the system in any positive integer spatial dimension.

It is worthwhile noting that quantum mechanics, considered as a quantum field theory in $d=0$, has no momentum integration in \eq{boinint} and yields finite Green's functions. Hence the singular structure of free Green's functions in quantum field theory with $d>0$ arises from the infinite volume of the external symmetry group.

The Wick rotation, a widely used method to rely on regulators in the Euclidean space-time, introduces some exponentially small but finite violation of the Lorentz symmetry, owing to the semi-circle contribution to the contour integral on the complex frequency space. It is rather natural that the Wick-rotation, the analytical continuation from the Euclidean space-time with the compact rotation group to Minkowski space-time with the non-compact Lorentz group is having difficulty in maintaining the manifest space-time symmetry. 

The problem with Lorentz symmetry has already surfaced under the disguise of a conditional UV singularity \cite{dyson} in the power counting argument for Minkowski space-time and its proposed solution, the use of a modified Feynman $i\epsilon$-prescription, $i\epsilon\to i\epsilon(\vv{p}^2+m^2)$ \cite{zimmermann}. But it is important to bear in mind that this is neither UV nor IR problem, it arises from the domain with large values of the coordinates or energy-momentum, used to parametrize the integration domain. Any finite value, assigned to \eq{boinint}, arises from the violation of the reparametrization invariance of the integral.

One can nevertheless arrive at a Lorentz invariant definition of loop-integrals with negative superficial degree of divergence \cite{itzykson} with a reasonable price. The point is that these loop-integrals are absolute and uniform convergent when $\epsilon\ne0$ and can be evaluated by the  usual procedure, by integrating over the components of the momentum in an arbitrary order according to Fubini's theorem. This freedom lends the appearance of being well defined to these integrals however this is not the case, the non-linear change of integral variables, like the use of invariants \eq{inv}, remains excluded. This problem, namely the definition of the theory by the help of the coordinate system with homogeneous metric tensor, is reminiscent of the well known loss of the symmetry of the quantization procedure under canonical transformation and the emergence of Ito potential in quantum mechanics \cite{schulman}. The acceptance of such a restriction of the definition of the loop-integrals leaves the regularization by higher order derivatives or by point splitting as possible relativistic cutoffs.

\section{Definite norm, stability, causality}\label{dscs}
We now turn to another feature of the regulator, the stability, mentioned together with two other equivalent conditions, within the context of free theories.

The cutoff dynamics should be stable however instabilities may arise from two sources: Higher than second order time derivatives in canonical equations of motion lead to Ostrogradsky's instability \cite{ostrogadsky} and runaway trajectories already in point particle mechanics. In field theories the cutoff hides the UV degrees of freedom which make up an environment for the regulated system, renders the cutoff theory open and the regulated action non-local. The energy of such theories is either unbounded from below or non-conserved \cite{nonloc}.

Another desired property is the preservation of the definite norm within the linear space of states. The higher order derivatives in the kinetic energy, the use of the propagator $G(\omega)=1/P(\omega^2)$ where $P(z)=\ord{z^{n_B}}$ is a polynomial with $n_B>1$, generates states with negative norm and invalidates the probabilistic interpretation of the quantum theory. In fact, the partial fraction representation of the propagator,
\be\label{pfr}
G(\omega)=\sum_n\frac{Z_n}{\omega^2-\Omega^2_n},
\ee
assuming single poles only, is $\ord{\omega^{-2}}$ unless some $Z_n$ are negative. 

The causality requires that the response to an external source appears after the action of the source and cutoff theories, being non-local, are usually do not possess this property. It is not entirely obvious whether the regulated theory should be causal at the cutoff since the dynamics is chosen artificially at this unobserved scale. Nevertheless a regularization procedure reflects our ignorance and is not supposed to introduce acausality, a qualitatively new and revolutionary feature. If it does then one has to assure that acausality is restricted to the cutoff scale. Even in that case such a regulator can be used in renormalizable theories only, excluding the more realistic class of effective theories. Acausality, if ever arises, should originate from a definite, observed phenomenon rather than from an auxiliary concept, such as the regulator. 

One can nevertheless make causality necessary by a more careful realization of the cutoff. The cutoff theory supports an open dynamics, owing to the unobserved UV degrees of freedom hence the regularization of quantum field theories is to be performed in a framework designed for open systems \cite{qrg}, namely within the Closed Time Path (CTP) formalism \cite{schw}. This is a CQCO scheme, it handles classical, quantum, closed and open systems on equal footing and treats initial rather than boundary value problems. The apparent redoubling of the degrees of freedom, the distinguishing feature of this scheme, allows the extension of the variational principle of classical mechanics for dissipative forces in open systems \cite{cctp} and the quantum effects arise as an $\cal O(\sqrt\hbar)$ separation of the two coordinates, describing the same degree of freedom. The additional advantage of this scheme is that it leaves the final state unrestricted thereby manifestly preserves the retardation nature of the interactions. The causality, a highly non-trivial feature of open, non-local dynamics appears in an explicit and controllable manner in the CTP scheme and becomes a necessary feature of the regularization.

The presence of states with indefinite norm, the Ostrogradsky's instability and acausality are equivalent for free fields: The sign of $n$ in the spectrum of a harmonic oscillator, $E_n=\hbar\omega(n+1/2)$, is the same as the sign of the norm of the excitation, generated by $a^\dagger$. The contribution with $Z_n<0$ in\eq{pfr} are associated with the propagation of particles with negative norm states. Finally, the instability can be ``hidden'' as an acausality in the retarded Green's function \cite{nonloc}.

\section{Point splitting}
The apparent incompatibility of the boost invariance and stability excludes the presently known regulators in relativistic free field theories and forces us to accept the usual formal generalized functions to define  the free Green's functions. However one can prevent the proliferation of the singularities in interactive theories by smearing the interactions in the space-time, called point splitting. The instability emerges here because the cutoff theory is non-local and its energy is unbounded from below \cite{nonloc}. The point splitting is a soft cutoff, it leaves arbitrarily high energy states in the Fock space and whatever small numerical effect the cutoff exerts on the motion the resulting instability modifies the dynamics at an arbitrary  scale. 

A point splitting regulator which satisfy all requirements is constructed in two steps. First the causal structure is established, followed by the steps, needed to assure stability.

\subsection{Causality}\label{pointspls}
Let us consider for the sake of simplicity a theory for a scalar field, $\phi(x)$, defined by the action $S=S_0+S_1$, 
\bea
S_0[\phi]&=&\hf\int dx\phi(x)D^{-1}\phi(x),\nn
S_1[\phi]&=&-\int dxU(\phi(x)),
\eea
where $D$ denotes the free propagator and $U(\phi)$ describes the self interaction. The point splitting consists of the use of the smeared, bare field variable given by the convolution 
\be
\phi_B(x)=\int dy\chi(x-y)\phi(y),
\ee
in the interaction,
\be
S_{1B}[\phi]=-\int dxU(\phi_B(x)).
\ee

The cutoff theory is open and should be defined within the CTP formalism. One reduplicates the degrees of freedom in this formalism and uses the CTP doublet field, $\hphi=(\phi^+,\phi^-)$ with the action
\bea\label{ctporig}
S_0[\hphi]&=&\hf\int dx\hphi(x)\hD^{-1}\hphi(x),\nn
S_1[\hphi]&=&-\int dx\left[U(\phi^+(x))-U(\phi^-(x))\right],
\eea
where the CTP propagator possesses the block structure
\be\label{grfnctstr}
\hD=\begin{pmatrix}D^n+iD^i&-D^f+iD^i\cr D^f+iD^i&-D^n+iD^i\end{pmatrix},
\ee
in terms of the near and far field Green function, $D^n$ and $D^f$, respectively \cite{ctpform}. The point splitted model is defined by the interaction,
\be\label{ctppspla}
S_{1B}[\hphi]=-\int dx\left[U(\phi_B^+(x))-U(\phi_B^-(x))\right],
\ee
written in terms of the bare field,
\be
\hphi_B(x)=\int dy\hat\chi(x-y)\hat\sigma\hphi(y),
\ee
where smearing is performed by a symmetric kernel, $\chi^{ab}(x-y)=\chi^{ba}(y-x)$ and $\hat\sigma=\mr{Diag}(1,-1)$. 

The particular way the interaction term is regulated allows us to rewrite the path integral expressions of the theory by performing the change of variable, $\hphi\to\hphi_B$, appearing in the action $\tilde S=\tilde S_0+\tilde S_1$, with
\be\label{highderke}
\tilde S_0[\hphi]=\hf\int dx\hphi(x)\hD_B^{-1}\hphi(x),
\ee
containing the regulated propagator $\hD_B=\hat\chi\hD\hat\chi$ and $\tilde S_1=S_1$, given by the original, local interaction in the second line of \eq{ctporig}. The Green's functions of $\hphi$ are obtained by multiplying each external leg of the Green's functions of $\hphi_B$ by $\hat\chi^{-1}$. The theory contains negative norm states of $\phi_B$ however their contribution is suppressed on the external leg by the factor $\chi^{-1}$ since the field $\phi$, having a traditional kinetic energy, generates states with positive definite norm only. 

Causality, a non-trivial property of point splitted theories, can be assured by using smearing a function, $\tilde\chi$, with the block structure \eq{grfnctstr} with $\chi^i=0$ to keep the bare field real. The retarded and advanced bare propagators, $\hD^{\stackrel{r}{a}}=D^n\pm D^f$, are given by $\hD_B^{\stackrel{r}{a}}=\chi^{\stackrel{r}{a}}D^{\stackrel{r}{a}}\chi^{\stackrel{r}{a}}$, where $\chi^{\stackrel{r}{a}}=\chi^n\pm\chi^f$. We restrict the smearing kernel by the equation  $\chi^f(x)=\mr{sign}(x^0)\chi^n(x)$ since it renders the bare dynamics causal, $\hD_B^{\stackrel{r}{a}}(x)\sim\Theta(\pm x^0)$.

A possible choice is 
\be\label{chidef}
\hat\chi=\sqrt{\prod_{j=1}^{N_c}(-c_j\Lambda^2)\mr{Re}\hD_{c_j\Lambda}}
\ee
with $c_i\ne c_j$ where $\hD_\Lambda$ is the CTP propagator of a scalar particle of mass $\Lambda$ and the constants $c_j$ are dimensionless regulator parameters. Note that the square root does not appear in the bare propagator owing to the commutativity of translation invariant operators. The real part, taken in the space-time, assures that the smeared Hermitian field remains Hermitian. The bare inverse propagator, $\hD_B^{-1}=\hat\chi^{-1}\hD^{-1}\hat\chi^{-1}=\hat\chi^{-2}\hD^{-1}$, contains higher order derivatives. The use of the higher order spatial derivatives hence leads to a stable, Galilean invariant cutoff theory. However higher order time derivatives appear in the relativistic case, considered in the rest of this work, where the resulting instability can be regarded as a generalization of Ostrogradsky's result. The square root in the definition \eq{chidef} and the non-degeneracy condition, $c_i\ne c_j$, are to avoid the multiple roots of the inverse propagator and the corresponding secular solutions of the equation of motion.  The bare propagator, defined in such a manner, has causal, real poles hence the free theory is stable however instability arises in the presence of interactions.

\subsection{Regularization in the internal space}\label{insspss}
An instability drives the classical fields to large values and generates highly occupied states in the quantum theory. Such a damage on the dynamics arises from the interactions alone and should be possible  to bring under control by suppressing the interactions for large field, by introducing a regulator in the internal space, for the field amplitude. A possible way to achieve this goal is the the use of the interaction 
\be
\tilde S_1[\hphi]=-\sum_aa\int d^{d+1}x\rho(\phi^a(x))U(\phi^a(x)),
\ee
for the bare field where $\rho(\phi)=0$ when $\phi\to\infty$. The instability, generated by the higher order derivatives, growths exponentially with the time hence $\rho(\phi)$ should decrease faster than any power of $\phi$, e.g.
\be\label{suppr}
\rho(\phi)=e^{-\frac\lambda{\Lambda^{2d_\phi}}\phi^2},
\ee
where $\lambda$ is a dimensionless parameter and $d_\phi=(d-1)/2$. In dimension $d=1$ where $\phi$ is dimensionless, $d_\phi=0$, one has to perform the limit $\lambda\to0$, as well.

The semi-classical stability can easily be verified numerically for the 1+1 dimensional theory, defined by the Lagrangian
\be
L=-\hf\phi(\Box+m^2)\left(1+\frac\Box{\Lambda^2}\right)\phi-\frac{g}{4!}e^{-\lambda\phi^2}\phi^4.
\ee
The theory is placed on a spatial lattice of sites $N$ and periodic boundary conditions to simplify the calculation. This is a system of interacting Pais-Uhlenbeck oscillators \cite{pu} and the non-interacting oscillators, $g=0$, are stable for non-degenerate spectrum, $m^2\ne\Lambda^2$, without secular solutions. The stability can be reconciled with Ostrogradsky's theorem by using an alternative Hamiltonians \cite{pumultiham}, and the interactive oscillator appears stable in certain cases \cite{pustab}. 

\begin{figure}
\includegraphics[scale=.5]{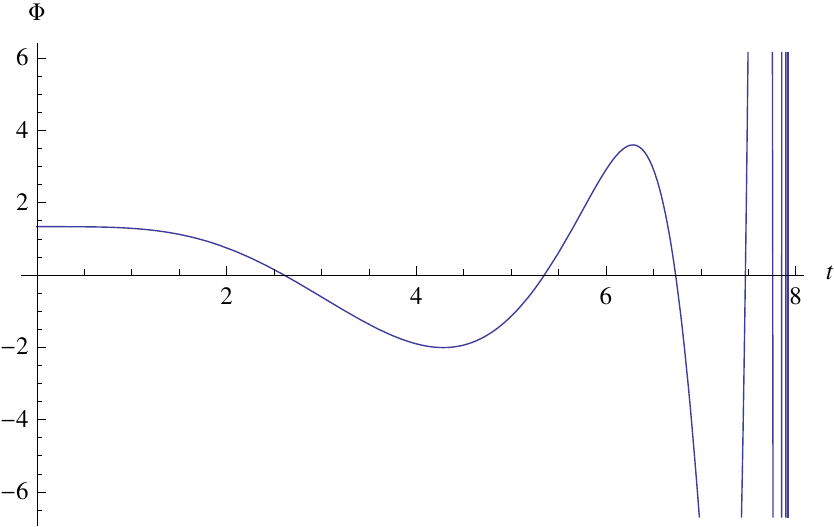}\hskip.5cm\includegraphics[scale=.5]{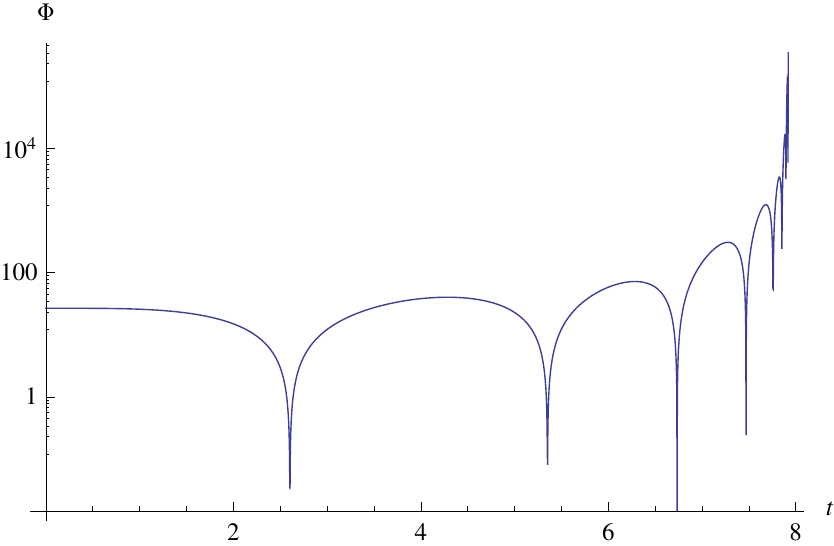}\hskip.5cm\includegraphics[scale=.5]{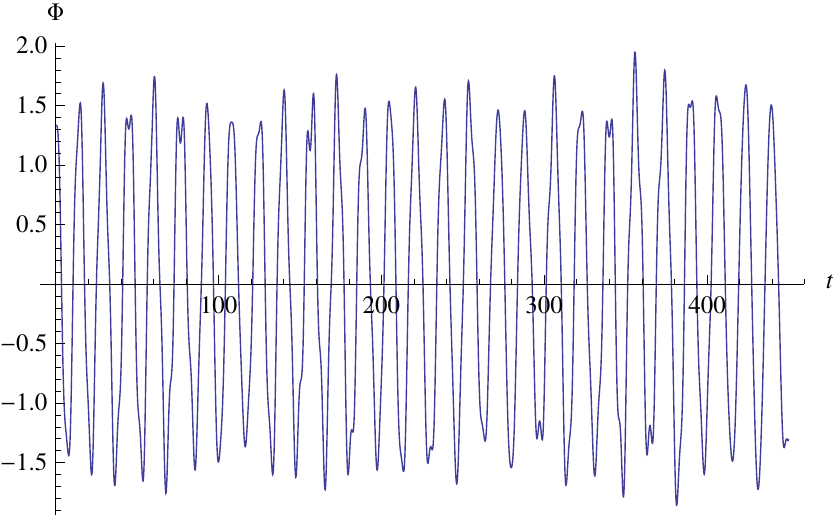}
\caption{The time dependence of the homogeneous mode. (a): $\Phi$, (b): $\ln|\Phi|$ with $\lambda=0$ and (c): $\Phi$ for $\lambda=1$.}\label{oscillf}
\end{figure}

The time evolution of the mode $p=0$, $\Phi=\sum_n\phi_n/N$, is shown for $N=20$, $a=1$, $m^2=0.1$, $\Lambda^2=2$, $g=1$ in Fig. \ref{oscillf}. The initial conditions are $\phi_n(0)=1+p_n$, $p_n$ being small fluctuations and $\partial_0^j\phi_n(0)=0$ with $j=1,2$ and 3. Figs. \ref{oscillf} (a) and (b) display the runaway trajectory for $\lambda=0$ without suppressing the interactions on linear and logarithmic plots. While the time scale of the oscillation shrinks and the equation becomes stiff as the divergence is approached the details of the trajectory remain independent of the numerical details of the integration. The suppressed interaction, $\lambda=0.4$, generates a stable trajectory, shown on Figs. \ref{oscillf} (c). 

The cutoff, introduced in the internal space, contains irrelevant operators and the corresponding coupling constants, the regulator parameter $\lambda$ in case of \eq{suppr}, should be made dimensionless by the help of the cutoff. The power counting detects no new divergences in this case and the perturbative renormalizability can be established as in the usual case, by expanding around a stable free theory.

\section{Conclusions}
The regularization of a field theory can be carried out on two different levels, either by rendering the free Green's functions finite or by regularizing only the interaction generated divergences. The boost invariance makes the momentum spectrum unbounded hence the non-perturbative regulators of the former class violate boost invariance. The latter regulator renders the dynamics non-local in time in relativistic theories and makes the stability of the dynamics an open question. It is shown numerically for the 1+1 dimensional scalar model that a combined application of cutoffs of the interactions both in the external and the internal spaces yields stable, boost invariant dynamics. 

The present work is actually a part of a particular project, the systematic derivation and treatment of the Abraham-Lorentz force in (quantum) electrodynamics. The usual problems about the runaway trajectories, generated by the radiation reaction force, are generated by the cutoff \cite{nonloc} and their solution requires a manifestly relativistic invariant, stable regulated electrodynamics. That goal can be reached in a similar manner to the method, presented here, by controlling the amplitude of some gauge invariant combination of the gauge field. The semi-classical stability is specially easy to guarantee by the suppression of the interaction on the light cone \cite{clelrad}.

The classical field theory when removed from its historical background responds to the no-go theorem of relativistic mechanics, stating the impossibility of describing interactive relativistic particles by means of local forces  \cite{relnogo}: The continuously distributed degrees of freedom of a field, equipped with the necessary retardation, provides a signal relay mechanism which satisfies the causal requirement of special relativity. The problems of the cutoff, discussed in this work, indicate that relativistic field theories have not yet completely solved the problem.

The hand-waving argument about the semi-classical stability leaves an important question open, namely whether one can actually prove the existence of stable, relativistic cutoff theories, either in the classical or quantum regime? In the absence of a proof we have to rely on numerical methods to define relativistic field theories. Such a limitation renders the axiomatic quantum field theory \cite{streater} where one assumes the Lorentz covariance of the renormalized, finite Green's functions, unjustified and leaves special relativity as an emergent phenomenon \cite{nielsen}.

\appendix
\section{Singularity of the free Feynman propagator}\label{greena}
While the divergence of the type \eq{boinint} concerns Green's functions with vanishing external momentum a similar divergence arises for Green's functions, expressed as functions of the space-time coordinates. In fact, since the volume of the momentum space region, corresponding to a fixed value of $F(p^2)$ is infinite we need a further regulator to render the integral with the oscillating $e^{-ipx}$ factor well defined. In particular the free Feynman propagator
\be\label{fprop}
D_F(x)=\int\frac{d^{d+1}p}{(2\pi)^{d+1}}\frac{e^{-ipx}}{p^2-m^2+i\epsilon},
\ee 
a distribution, given in terms of a non-absolute convergent integral, needs a regulator for $d>0$. We present briefly three different ways to achieve this goal. Since the UV regularization is independent of the mass we consider the massless case for the sake of simplicity.

(i) The most natural way to deal with the Fourier integral is to carry out the energy integral first by the help of the residue theorem. The integration over the direction of the momentum leaves an integral over the length of the three-momentum,
\be
D_F(x)=-\frac1{8\pi^2|\vv{x}|}\int_0^\infty dk(e^{ik(|\vv{x}|-|x^0|)}-e^{-ik(|\vv{x}|+|x^0|)}),
\ee
when $d=3$. The quadratic divergence of \eq{fprop}, indicated by the power counting, is reduced to a linear one by the factor $1/|\vv{x}|$ but is still strong enough to make this integral ill-defined. It is customary to introduce an infinitesimal parameter, $\eta$, at this step,
\be
D_F(x)=\int\frac{d^3k}{(2\pi)^3}e^{i\vv{k}\vv{x}-\eta|\vv{k}|}\int\frac{dk^0}{2\pi}\frac{e^{-ik^0x^0}}{k^{02}-\vv{k}^2+i\epsilon},
\ee
which renders the propagator well-defined but leads to,
\be
D_F(x)=\frac{i}{4\pi^2x^2-2i|x^0|\eta}
\ee
by violating Lorentz invariance. This latter can be recovered by performing the limit $\eta\to0$, yielding
\be
D_F(x)\to P\frac{i}{4\pi^2x^2}-\frac1{4\pi}\delta(x^2).
\ee

(ii) The proper time method is based on the representation
\be
\frac1{k^2+i\epsilon}=-i\int_0^\infty dse^{is(k^2+i\epsilon)},
\ee
of the propagator in the momentum space, giving
\be
D_F(x)=-i\int\frac{d^4k}{(2\pi)^4}\int_0^\infty dse^{is(k^2+i\epsilon)-ikx}.
\ee
The Fresnel integral is ill defined and needs a further regulator, 
\be
\int\frac{d^4k}{(2\pi)^4}e^{isk^2-ikx}\to\int\frac{d^4k}{(2\pi)^4}e^{\sum_\mu g_{\mu\mu}[(is-g_{\mu\mu}\eta)k^{\mu2}-ik^\mu x^\mu]}.
\ee
This regulator obviously violates boost invariance,
\be
D_F(x)=-\frac1{32\pi^2}\int_0^\infty\frac{ds}{s^2}e^{-s\epsilon-\frac{i}{4s}(x^{02}-\vv{x}^2)-\frac\eta{4s^2}(x^{02}+\vv{x}^2)}\left[1+\ord{\left(\frac\eta{s}\right)^2}\right],
\ee
however the symmetry is recovered as $\eta\to0$.

(iii) The higher order derivative scheme, described in Section \ref{pointspls}, leads to the Feynman propagator,
\be\label{pwms}
D_B(p)=\frac1{p^2+i\epsilon}P\prod_{j=1}^{N_c}\frac{c_j\Lambda^2}{c_j\Lambda^2-p^2},
\ee
for the smeared field. The standard integration over the energy by means of the residue theorem, followed by the integration over the three-momentum produces a finite result and the propagator
\be\label{pwst}
D_F(x)=\prod_{j=1}^{N_c}\left(1+\frac{\Box}{c_j\Lambda^2}\right)D_B(x)
\ee
for the original field of the model regains the singularities of the free propagator without violating the Lorentz symmetry. 

While the regulator, the parameter $\eta$, is explicit in the cases (i) and (ii) the regularization with higher order derivatives preserves the boost symmetry as long as the restriction, mentioned at the end of section \ref{symms}, the use of the energy and momentum as integral variables, is accepted.

\end{document}